\documentclass{aa}  
\usepackage{xcolor}
\usepackage[hidelinks,colorlinks=true,linkcolor=blue,citecolor=blue]{hyperref}
\usepackage{cleveref}

\usepackage{graphicx}
\usepackage{array}
\usepackage{makecell}
\usepackage{txfonts}

\begin{document}

   \title{Cutting with precision}

   \subtitle{Leveraging Collapse Volumes to generate the next generation of zoom-in initial conditions}

   \author{B. A. Seidel\inst{1},
           K. Dolag\inst{1,2}
          \and
           J.G. Sorce\inst{3,4}
          }
   \authorrunning{Seidel et al.}

   \institute{Universit\"ats-Sternwarte, Fakult\"at f\"ur  Physik, Ludwig-Maximilians Universität, Scheinerstr. 1, 81679 M\"unchen, Germany\\
   \email{bseidel@usm.uni-muenchen.de}
         \and
         Max-Planck-Institut für Astrophysik, Karl-Schwarzschild-Straße 1, 85741 Garching, Germany
         \and
         Univ. Lille, CNRS, Centrale Lille, UMR 9189 CRIStAL, 59000 Lille, France
          \and
          Université Paris-Saclay, CNRS, Institut d’Astrophysique Spatiale, 91405 Orsay, France\
             }

   \date{Received September 15, 1996; accepted March 16, 1997}

% \abstract{}{}{}{}{} 
% 5 {} token are mandatory
 
  \abstract
   {}
   {Astrophysical processes happen across a wide range of spatial and temporal scales. This poses a significant challenge from the perspective of modeling the Universe and its evolution. Modern cosmological simulations attempt to square the circle of maximizing simulation volume in order to capture the full range of the density power spectrum while simultaneously optimizing spatial resolution for improved modeling of dynamics at the galaxy scale and below. Performing high resolution zoom-in simulations of galaxy clusters is a way to reconcile computational cost, mass resolution, and large scale realism in particle based numerical simulations. In order to study the baryonic evolution of structures it is critical to ensure that the volume of interest is uncontaminated by  high-mass low-resolution particles from outside. We introduce a new method of constructing stable boundaries for layered zoom-in initial conditions. Applying this method to massive galaxy clusters from the SLOW constrained simulation of the local Universe, we introduce the SLOW cluster (SLOW-C) zoom-in initial conditions.}
   {We select our target regions using a forward run of an intermediate-resolution gravity-only version of the parent box. We find the zero-velocity surface in the far future and define the Lagrangian region within it as the high resolution volume. To soften the effect of the resolution decrease at the boundaries, we employ a self-similar multi-mass layering scheme with up to four intermediate layers of decreasing resolution enveloping the high-resolution region.}
   {We performed test simulations for a first set of 20 such regions created from the SLOW constrained simulations containing 30 of the local galaxy clusters. The simulations at $z=0$ demonstrate these initial conditions to be remarkably stable against deformation and mixing of the boundary region. This hold at every point in time during the simulation run. Consequently they are uncontaminated to an unprecedented degree, reaching pristine regions of sizes consistently exceeding $6r_\mathrm{vir}$ and minimizing numerical problems often appearing in the contaminated part of the simulation volume.}
    {These simulations will provide the basis for the first, very high-resolution simulations of a large set of 
    %directly comparable 
    local galaxy cluster analogues and their environment to date. Their high fidelity in terms of boundary stability and resolution in combination with the accuracy of the underlying local Universe model makes these simulations the first of their kind. They will enable comparisons with highly resolved state-of-the art observations targeting both cluster properties and ICM physics as well as galaxy evolution in the local Universe.}

   \maketitle
%
%-------------------------------------------------------------------
\section{Introduction}
One of the biggest challenges in studying the Universe is the multi-scalar nature of physical processes in nature. In astrophysics, dynamics on Megaparsec to Gigaparsec scales can be non-trivially connected to processes on Parsec to Kiloparsec scales. Large cosmological numerical simulations including hydrodynamics are especially suited to study structures and processes on large and small scales simultaneously. There is, however a computational tradeoff that needs to be considered: Ideally one wants to simulate a large enough region to capture the largest actively collapsing density perturbations - or there is a risk of large structures being affected by boundary effects or simply not forming at all \citep[e.g.][]{bagla2005}. On the other hand, the spatial resolution needs to be high enough to allow to follow hydrodynamics. An even higher mass resolution is needed, when galaxy formation physics has to be followed \citep{1997MNRAS.288..545S}, or the turbulent dynamo process in the intra-cluster medium (ICM) has to be resolved within a magneto-hydrodynamical framework. \citep{2024ApJ...967..125S}.  
%even when studying if hydrodynamical effects to capture the smaller scale physical processes in sufficient detail \citep[e.g.][]{kimmig2025c}. 
This has become ever more important in the age of large-scale galaxy surveys like \textit{Euclid} and LSST, which can probe galaxies down to stellar masses as low as $10^8M_\odot$ \citep{euclidcollaboration2025a,brough2020}. For this reason current cosmological simulations have to skirt the boundary of modern high performance computing.

\begin{table*}[h!]
\tiny
\centerline{
\scalebox{.8}{
\begin{tabular}{|c|c|c|c|c|c|c|c|c|c|c|c|c|c|c|}
\hline
    & BS1996 & E1998  & Hutt & B2006 & Dianoga & The300 & C-EAGLE & Rom-C & TNG-C & PICO &Rhapsody-G&MACSIS&Virgo CLONE &SLOW-C\\
\hline
    $N_\mathrm{cluster}$ & 10 & 10 & 8 & 3& 27 & 324 & 30 & 1& 325 & 25 &10 &390&1&45\\
\hline
    \makecell{$M_{\mathrm {200c,max}}$ \\ $[10^{15}h^{-1}M_\odot]$} & 2.4 & 1.8 & 2.3 & 1.3 & 3.2 & 2.6 &2.5 &0.2&2.5 & $\sim2.5$ &1.3&$\sim $3&$\sim$ 0.4&1.5 \\
\hline
    \makecell{$L_\mathrm{parent}$ \\ $[h^{-1}{\rm cMpc}]$} & 150 & 180 & 479 & 192 & 1000 & 1000 &2169&34& 680 & 1000 &1000 &2200&500&500 \\
\hline
    \makecell{$R_\mathrm{clean}$ \\ $[R_\mathrm{vir}]$} & 0.1-2.6 & 4-8 & 3.7-8.1 & - & 5.0-6.0 & 0.1-7 &-& - & 0.15-5.0 & 1.1-4.8 &-&-&$\approx 5$& $\approx10$ \\
    
\hline
    \makecell{$m_\mathrm{DM}$ \\ $[h^{-1}M_\odot]$} & $3.2\times10^{11}$ & $1.8\times10^{10}$ & $2\times10^{9}$ & $4.8\times10^{8}$ & $3.3\times10^6$ & $1.5\times10^9$ &$6.6\times10^6$& $2.3\times10^5$ & $4.1\times10^7$ &$4.0\times10^7$&$1.3\times10^8$&$4.4\times10^9$&$3\times10^7$& $3.9\times10^7$ \\
\hline
\end{tabular}}}
    \caption{Overview of the characteristic of different zoomed cluster simulations campaigns. From left to right BS1996 \citep{1996MNRAS.283..431B}, Eke11998 \citep{1998ApJ...503..569E}, Hutt \citep{2004A&A...416..853D}, B2006 \citep{2006MNRAS.367.1641B}, Dianoga \citep{bonafede2011}, The300 \citep{cui2018}, C-EAGLE \citep{barnes2017}, Romulus-C \citep{tremmel2019}, TNG-Cluster \citep{nelson2024a}, PICO \citep{tevlin2025} \textsc{Rhapsody-G} \citep{wu2015}, MACSIS \citep{barnes2017b}, the related Virgo \textsc{Clone} \citep{sorce2021} and SLOW-Cluster (this work). Listed are the number of clusters in the set, the size of the parent box, the size of the cleaned region, e.g. the (typical) distance from the center to the clusters to the nearest boundary particles at $z=0$ (where available) and the mass of the dark matter mass particles at the highest resolution available for the set. Here we focus exclusively on $\Lambda$CDM models, there are several projects using non-$\Lambda$CDM initial conditions \citep[e.g.][]{nadler2025a,an2025} as well as self-interacting dark matter models \citep[e.g.][]{nadler2025,harvey2025}}
\label{sim:props}
\end{table*}

One solution for this challenge are so called zoom-in techniques \citep[e.g.][]{frenk1996,tormen1997,1999A&A...348..351D,2004A&A...416..853D,jenkins2010,bonafede2011,hahn2011b} that are widely applied to particle-based simulations as well as hydrodynamical grid codes. Here, the resolution of the simulation is increased in a specific Lagrangian volume of interest, with the rest of the pseudo-particles simulated at a lower resolution to save computational cost. A major challenge of this approach is to choose the Lagrangian region in such a way that low-resolution and high-resolution particles mix as little as possible for the duration of the simulation. This mixing can introduce spurious energy transfers due to particles of different masses interacting. This is especially an issue in the high-resolution region, where high-mass boundary particles would not only interact with high-resolution dark matter particles but additionally the collisional baryonic particles and stellar particles in SPH simulations. These baryonic particles usually have even smaller masses (in our Magneticum-like approach this is simply the dark matter mass multiplied by the cosmological baryon fraction).

Because of this, a major issue of modern particle based multi-mass zoom-in simulations is contamination - the intrusion of high-mass low-resolution boundary particles deep into the high-resolution region.  \citep{onorbe2014a} demonstrated that while the effect of contamination on purely gravitational dark matter dynamics is weak in terms of general halo properties, even a small amount of boundary particles can alter sensitive baryonic properties. They showed the gas density profiles and baryon fractions to be significantly altered in haloes with contamination versus the uncontaminated case, with an overall effect of a decreased baryonic density in the contaminated halos. The authors furthermore showed that the high-mass intruders can act as sink particle for the gas creating additional non-physical substructures. As the aim of zoom-in simulations is precisely to resolve the small scale gas dynamics in greater detail than pure cosmological simulations, it is therefore critical to keep the contamination as low as possible. In this work we propose a selection method based on extrapolated bound regions of galaxy clusters in order to achieve this. In \cref{sec:methods} we will discuss our approach in detail.

This paper is structured as follows: In section 2 we describe historical and current zoom-in simulations and how their initial conditions (ICs hereafter) were derived as a direct reference for this work. In section 3 we introduce our method in detail. In section 4 we compare the results from our simulations to the previously described reference simulations. Section 5 provides a summary and final discussion of our findings.

\section{A brief history of cluster zoom-in simulations}

\begin{figure*}[ht!]
   \centering
   \includegraphics[width=0.195\textwidth]{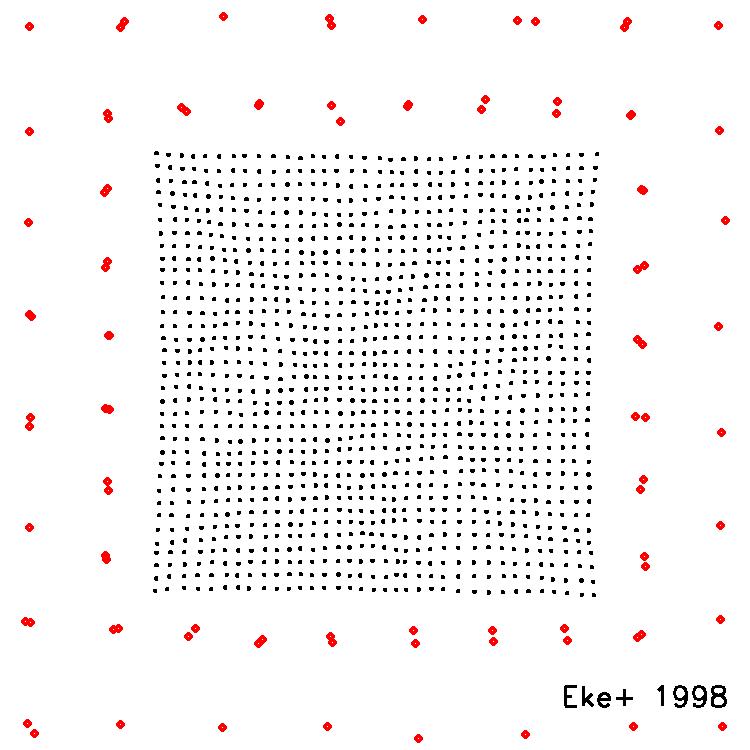}
   \includegraphics[width=0.195\textwidth]{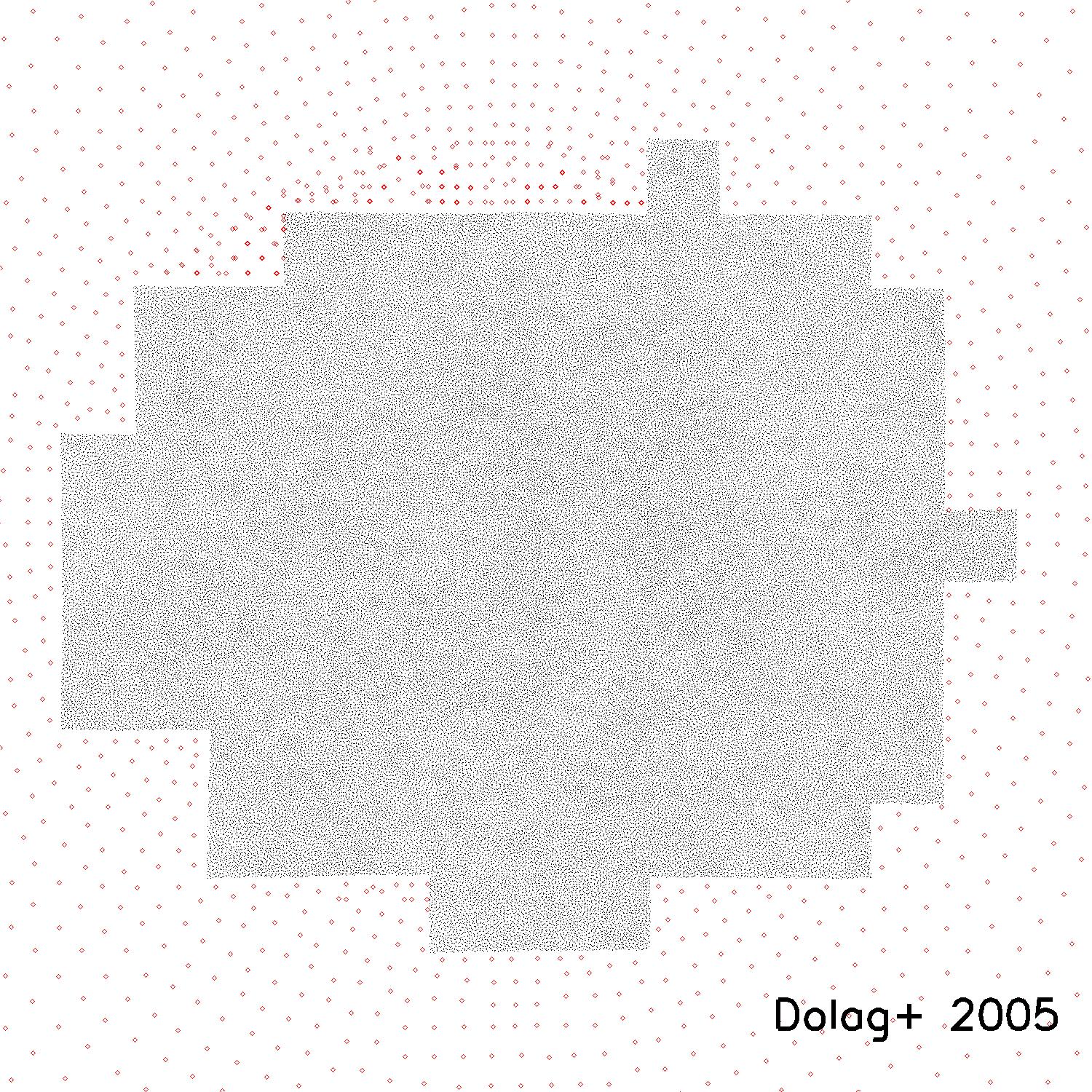}
   \includegraphics[width=0.195\textwidth]{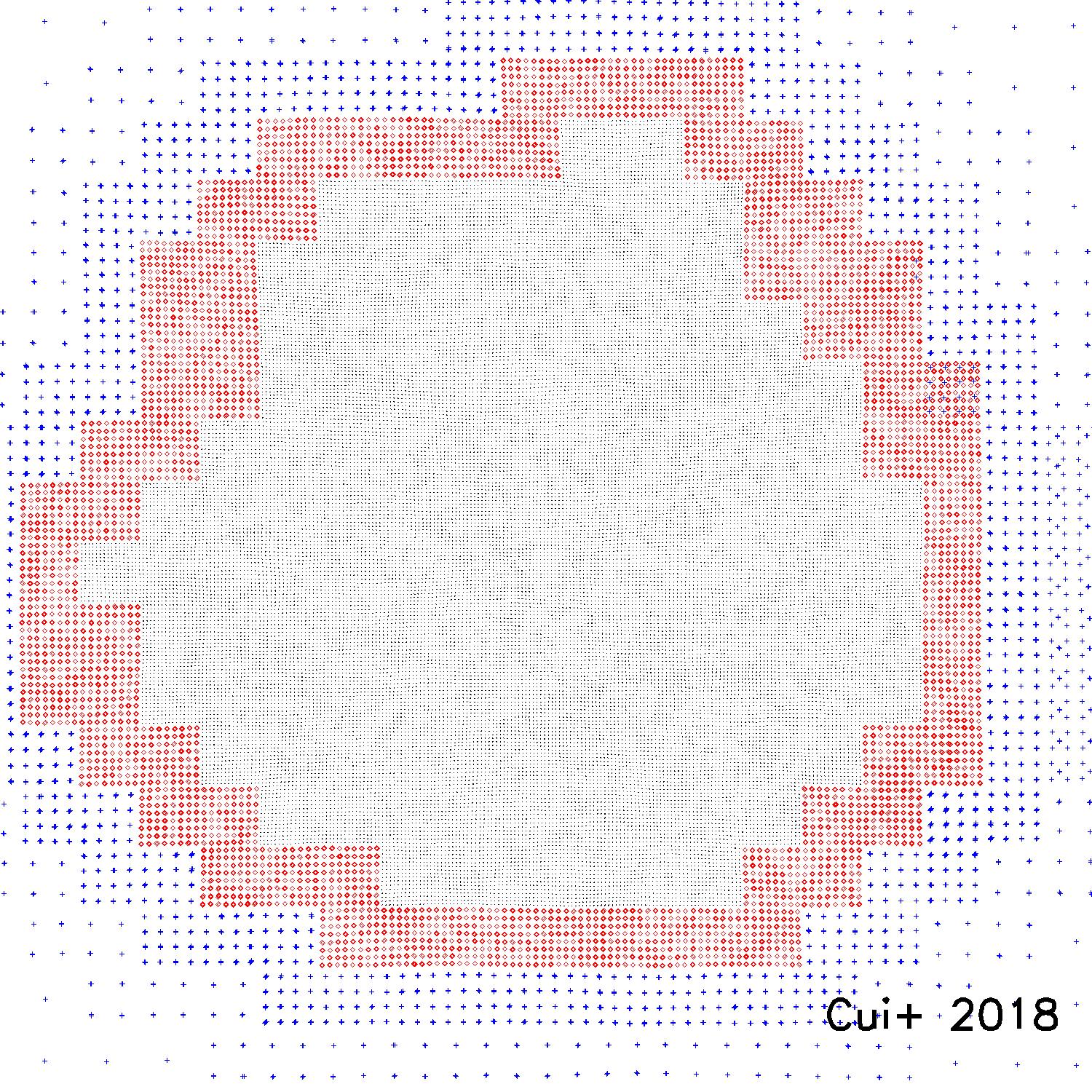}
   \includegraphics[width=0.195\textwidth]{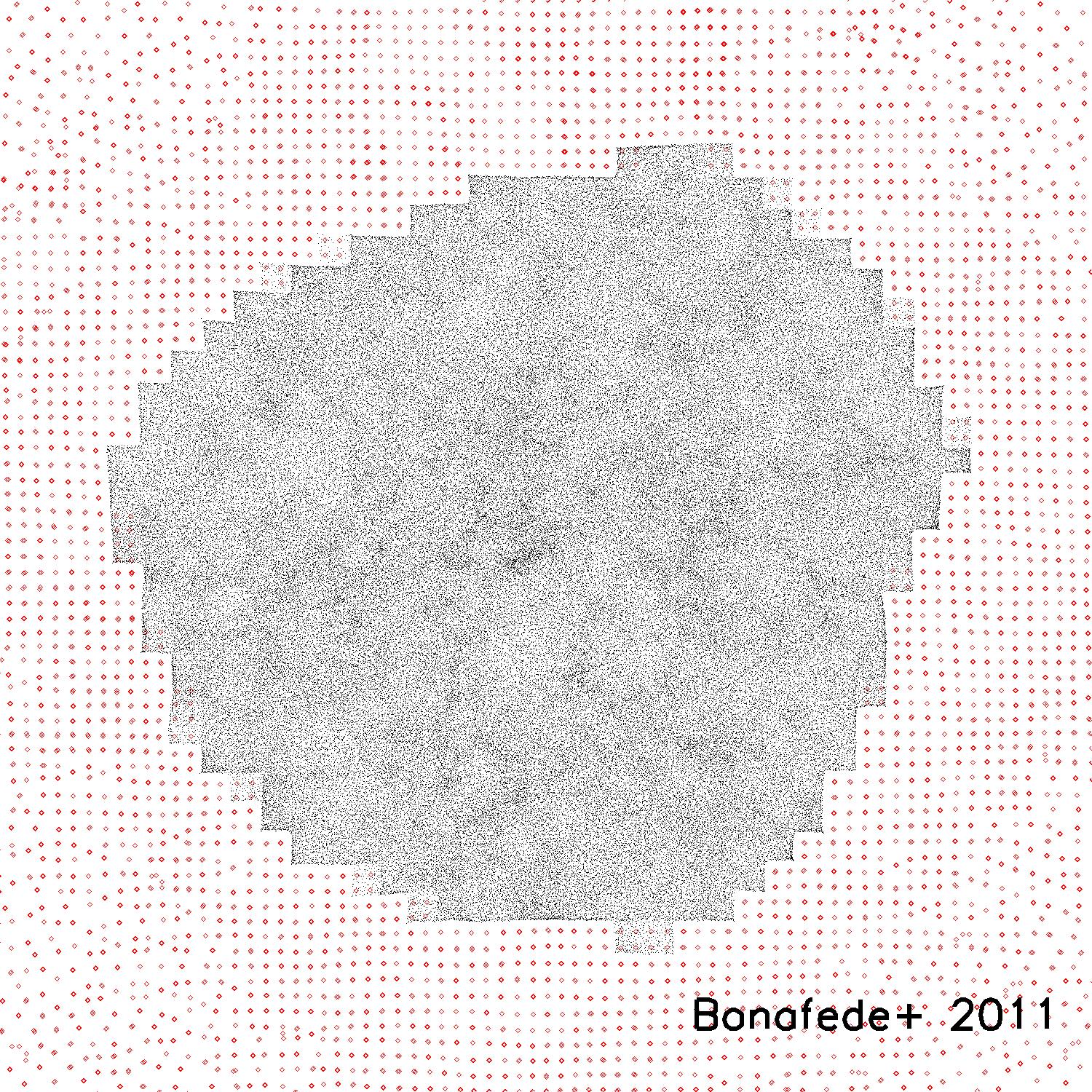}
   \includegraphics[width=0.195\textwidth]{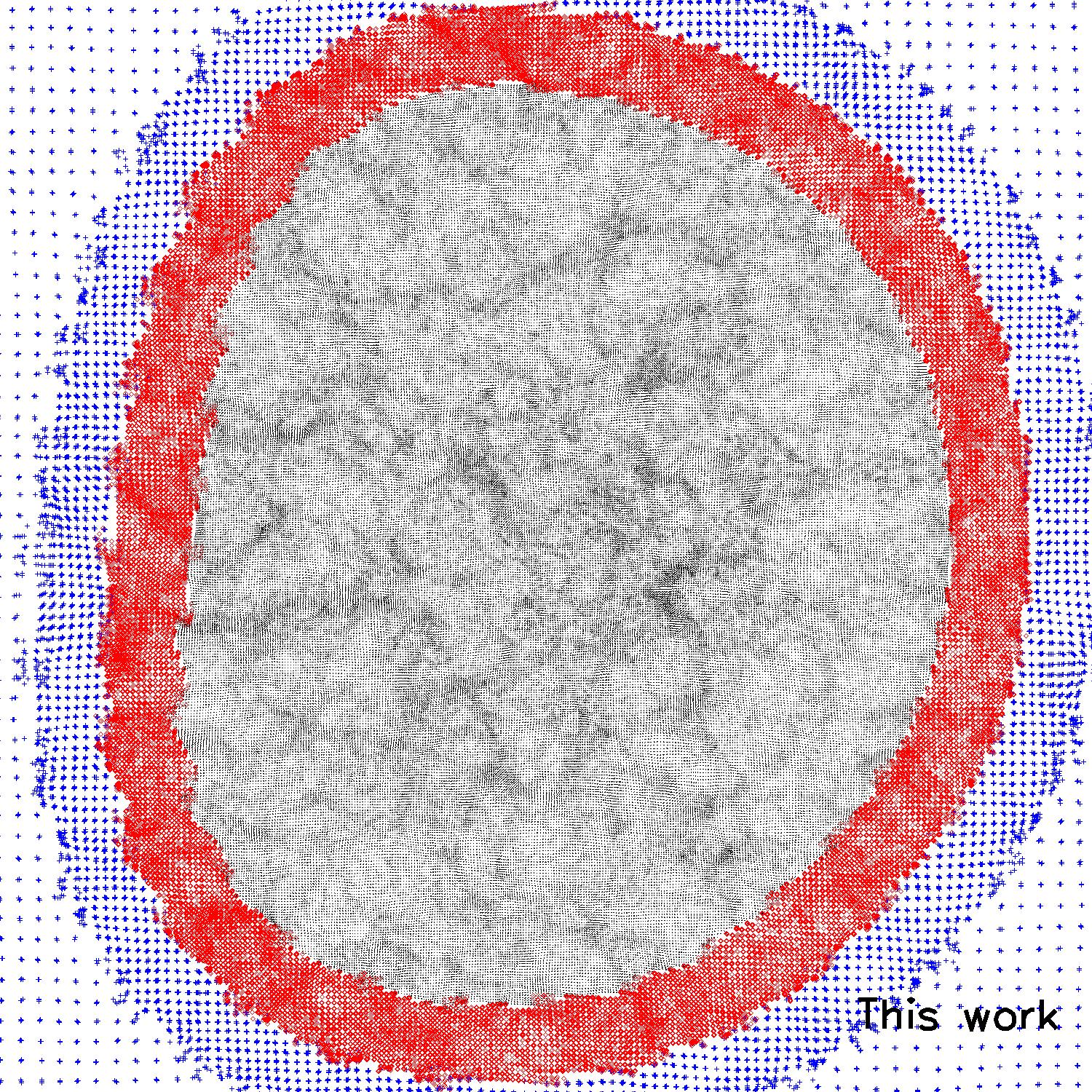} \\
   \includegraphics[width=0.195\textwidth]{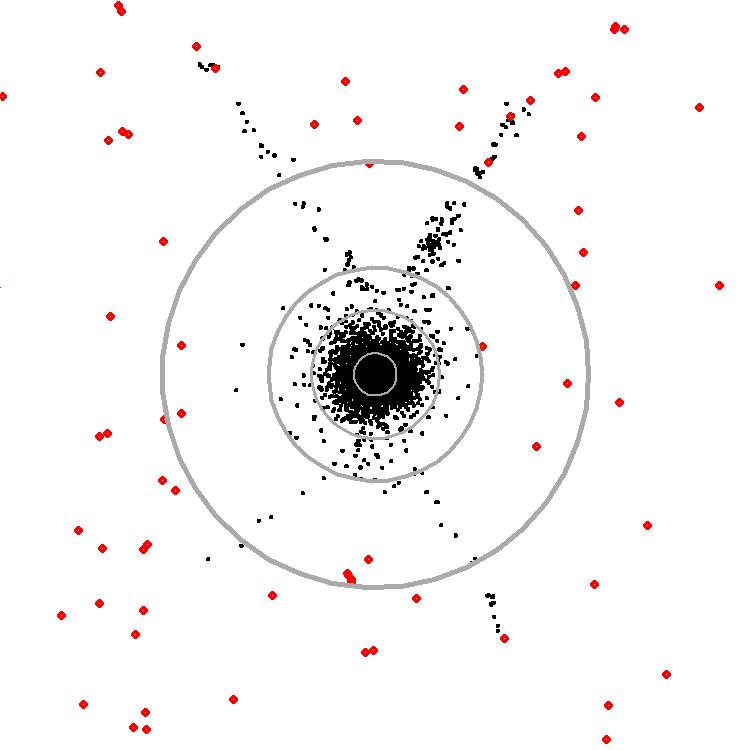}
   \includegraphics[width=0.195\textwidth]{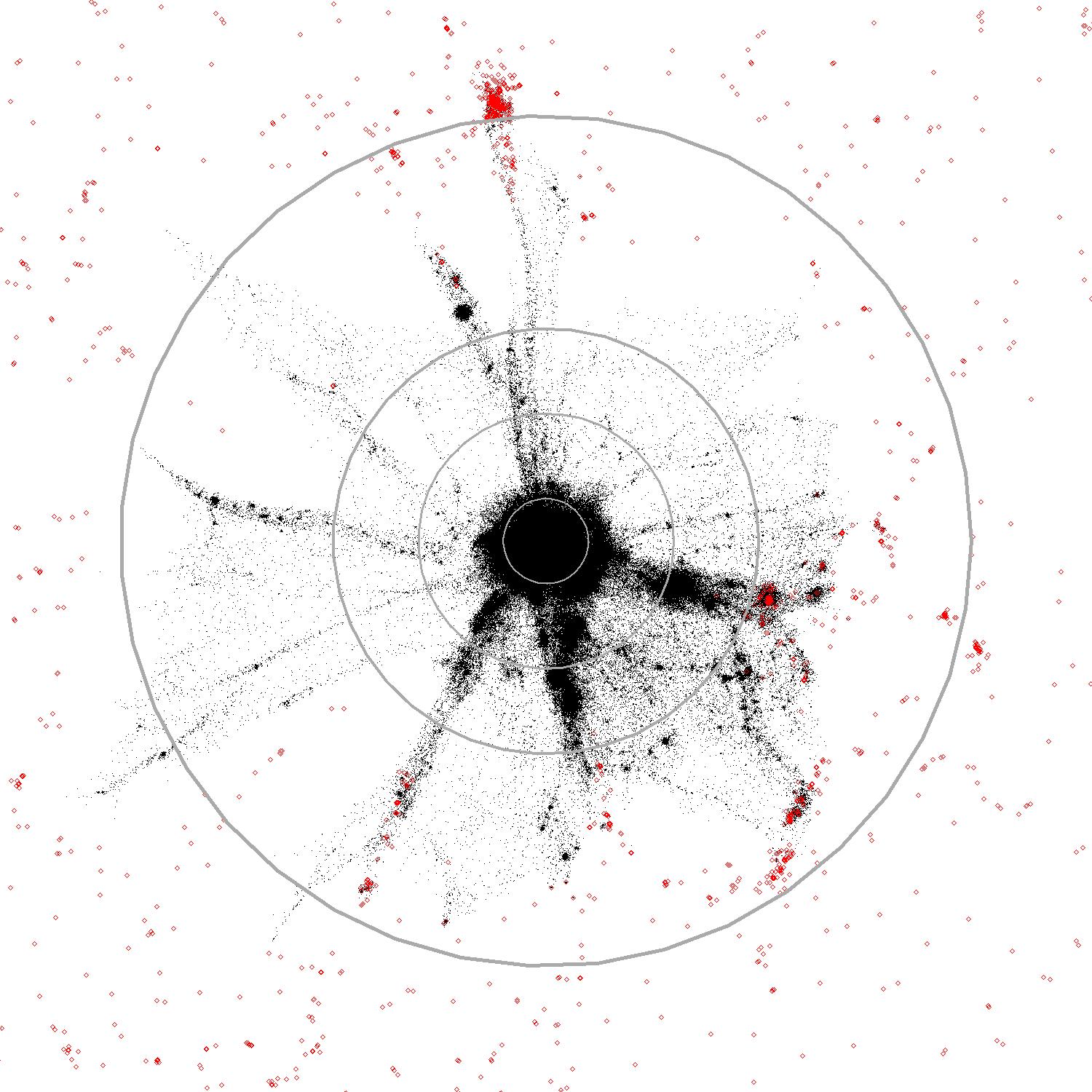}
   \includegraphics[width=0.195\textwidth]{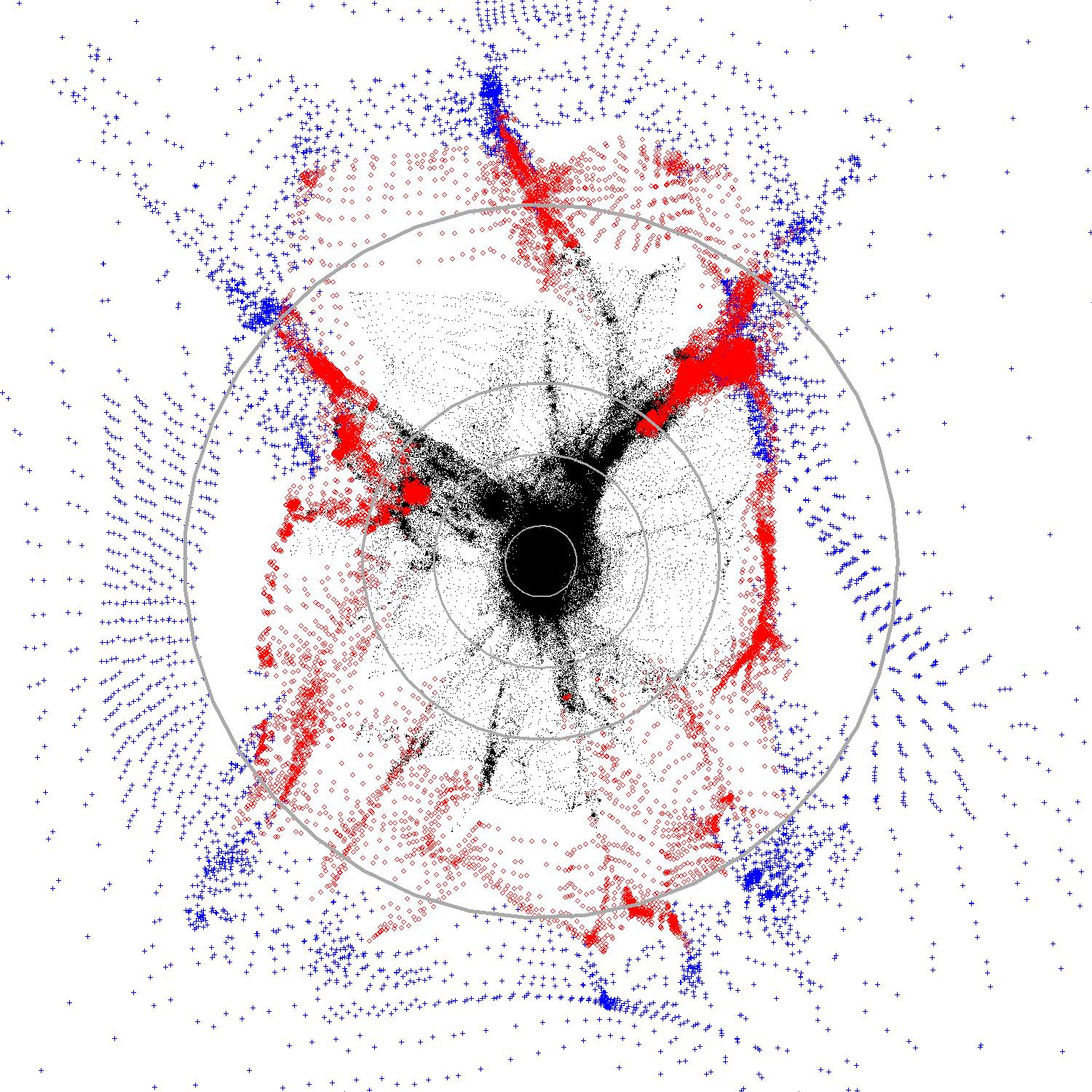}
   \includegraphics[width=0.195\textwidth]{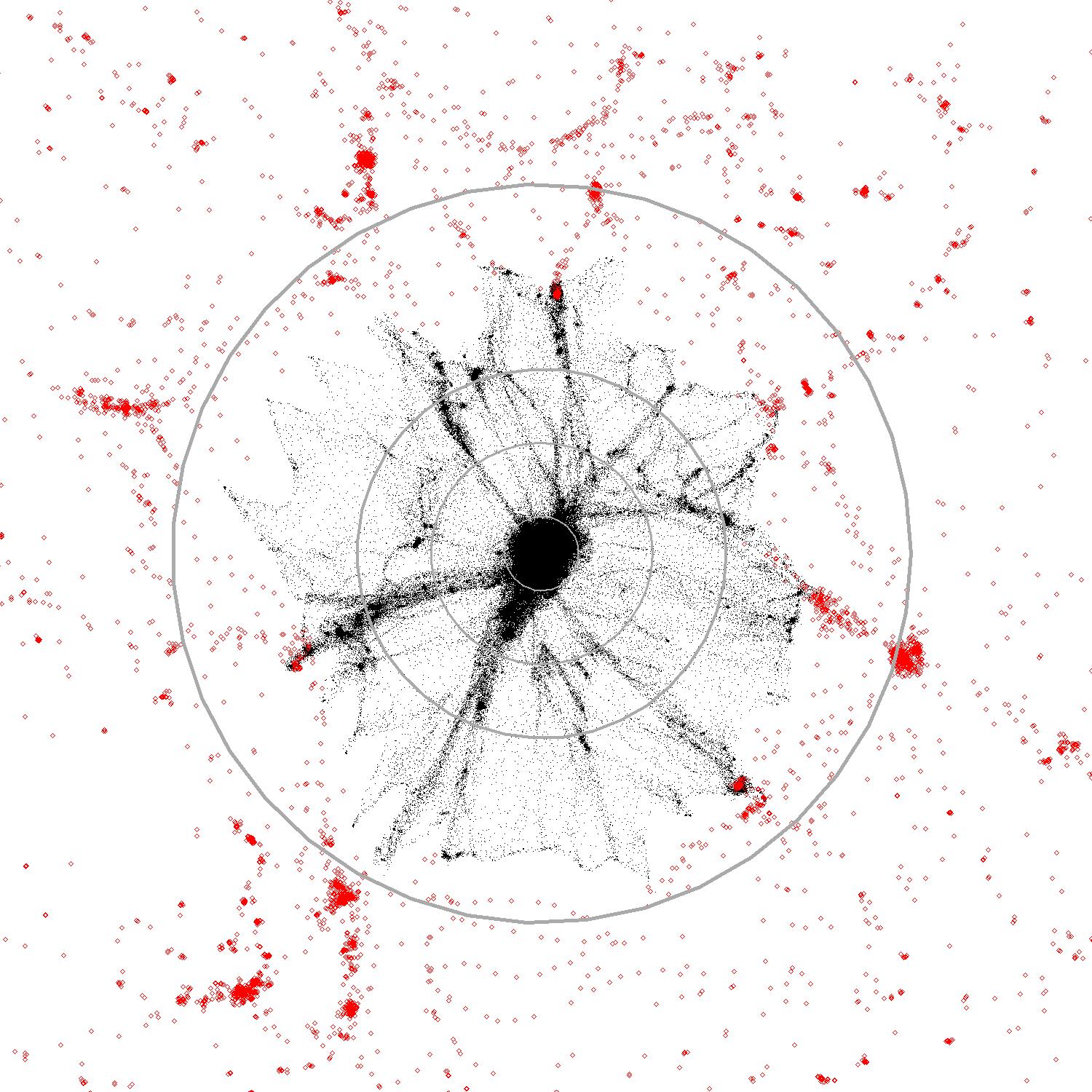}
   \includegraphics[width=0.195\textwidth]{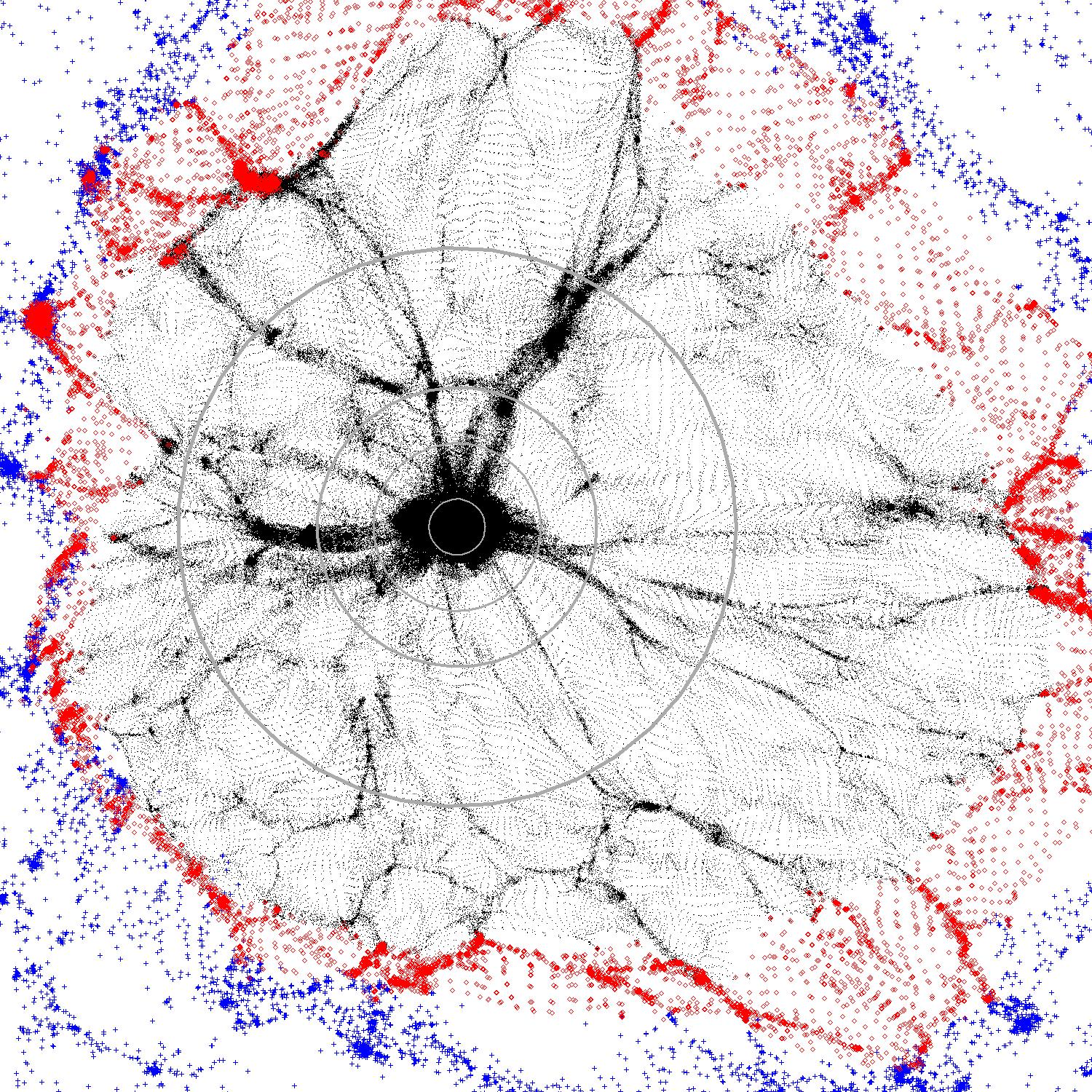}

   \caption{Thin slice of one cMpc/h through galaxy clusters from different simulation sets, showing a 70 cMpc/h region. For different sets of cluster ICs we show the high resolution particles (black) and the different layers of boundary particles (red and blue). The top row shows the initial state, the bottom row shows the evolved systems at $z=0$. From left to right: on cluster from  Eke1998 \citep{1998ApJ...503..569E}, Hutt \citep{2004A&A...416..853D}, Nifty (e.g. The 300) \citep{cui2018}, Dianoga \citep{bonafede2011} and SLOW cluster set (this work). The gray circles mark 1,3,5 and 10 times the virial radius. }
              \label{sim:overview}
\end{figure*}

From the very beginnings of galaxy cluster simulations it became clear that, in order to model the ICM physics with enough detail to predict observational properties of clusters, it is necessary to boost the resolution of cosmological volume simulations in the regions where those clusters form. The first such cluster zoom-in ICs were created by running a low resolution parent box to $z=0$, identifying over-densities corresponding to massive clusters, tracing them back to the ICs and resampling cubic regions around this volume with higher resolution particles. Table \ref{sim:props} lists a selection of sets for different zoom-in cluster ICs together with their properties. The first sets of this kind were employed to, for example, study the impact of sub-structures on the strong lensing signal \citep{1995A&A...297....1B} and predictions of the evolution of X-ray scaling relations for clusters \citep{1998ApJ...503..569E}. These sets typically contained ten clusters with masses of $\approx10^{15}h^{-1}M_\odot$ extracted from parent simulations with sizes of $\approx 300$Mpc. Those where also used to study the evolution of magnetic fields \citep{1999A&A...348..351D} as well as to address biases for different mass estimates for galaxy clusters \citep{1996MNRAS.283..431B,2000A&A...364..491D}.    

With the introduction of ZIC \citep{tormen1997} a significant advance was made, as here the region which was replaced with high resolution particles could be described by a cubic lattice, based on $16^3$, $32^3$ or $64^3$ cells which allowed to extract high-resolution volumes of arbitrary shapes. It also introduced an iterative procedure, where low resolution zoom-in simulations were used to refine the selected region. This made it possible to have the final cluster free of contaminating boundary particles while minimizing the required volume. Another advantage of ZIC is that it allows to place an arbitrary number of high resolution particles within the zoom-in region and the tool therefore does not depend on an underlying mesh where spatial resolution can only be increased by factors of two. This was originally used to extract a small set of massive clusters with masses of $10^{15}h^{-1}M_\odot$ \citep[Hutt,][]{2004A&A...416..853D} from a 479 $h^{-1}$Mpc box \citep{2001MNRAS.328..669Y} as well as creating three extremely high resolution clusters \citep[B2006,][]{2006MNRAS.367.1641B} from a 192 $h^{-1}$Mpc, full hydrodynamical simulation \citep{2004MNRAS.348.1078B}. Later the same technique was applied to a 1 $h^{-1}Gpc$ box to extract the Dianoga set of 24 massive (e.g. $M>10^{15}h^{-1}M_\odot$), Coma like clusters \citep{bonafede2011}. 

These sets of ICs were especially created to have large contamination free regions, e.g. $R_\mathrm{clean}>3.5R_\mathrm{vir}$ (Hutt) and $R_\mathrm{clean}>5R_\mathrm{vir}$ (Dianoga). Among many other studies, they have been used to study the thermodynamics of the ICM \citep{2011MNRAS.416..801F}, the role of AGN feedback for galaxy clusters \citep{2014MNRAS.438..195P}, the chemical enrichment of the ICM \citep{2017MNRAS.468..531B} and intra-cluster light properties \citep{dolag2010}. Recently, the resolution of these ICs have been further increased and reached levels of $10^9$ particles within individual clusters, allowing to fully resolve the turbulent dynamo within the ICM \citep{2024ApJ...967..125S}.

The three-hundred (The300) project \citep{cui2018} consists of 324 galaxy clusters extracted from a 1 $h^{-1}$Gpc parent box using MUSIC \citep{Hahn_2011}. MUSIC also allows for the definition of arbitrary high-resolution volumes, but the spatial resolution can be only increased by factors of two. Additionally, no iterative cleaning procedure was performed by the authors. Therefore, as listed in table \ref{sim:props}, there are clusters where boundary particles are present within the virial radius. The set was originally 
simulated employing a number of different subgrid physics models such as \textsc{Gadget-X} \citep{springel2005h,beck2016e}, \textsc{Gadget-music} \citep{sembolini2013} and a selection of semi-analytical models for galaxy formation. Recently, this was extended to include a GIZMO-SIMBA \citep{2022MNRAS.514..977C} variant as well as different hydrodynamical models. Studies carried out with this suite include, but are not limited to exploring the connection between the dynamical state and the morphology of these clusters \citep{deluca2021}, filament detection studies \citep{kuchner2021b,cornwell2023} and environment-galaxy connection studies \citep{wang2018}.

An even larger parent box of $2169h^{-1}$Mpc in size is used in the C-EAGLE project \citep{barnes2017}. This set consist of 30 zoom in galaxy clusters in the mass range from $10^{14}-10^{15.4}M_\odot$, that include as well the Hydrangea \citep{bahe2017} set of clusters simulated specifically to study galaxy formation. Simulations were run with the EAGLE model \citep{schaye2015b} with a modified \textsc{P-Gadget3} from \citep{springel2005h}. The simulations were used to study the infall of gas onto galaxy clusters through filaments \citep{vurm2023}, the galaxy mass functions in those regions \citep{negri2022} as well as the disruption of infalling satellite galaxies  \citep{bahe2019}.

To push the resolution to much higher level, the Romulus-C project \citep{tremmel2019} features a low mass galaxy cluster with $2\times10^{14}h^{-1}M_\odot$. It is selected from a $34h^{-1}$Mpc parent box from the Romulus cosmological simulations \citep{tremmel2017} and run with an identical setup in terms of hydrodynamics and sub-grid modeling. The project, due to its high mass-resolution, focused mainly on member galaxies and specifically dwarf galaxies, investigating environmental quenching and thermal AGN feedback effects \citep{tremmel2017} as well the formation of ultra-diffuse galaxies in cluster environments \citep{tremmel2020}.

The TNG-Cluster project \citep{nelson2024a} pushed for a large set of clusters with higher resolution than the The300 set. It is a suite of 325 galaxy clusters selected from a $680h^{-1}$ Mpc parent box and therefore has a similar mass range than previous projects and was ran with the TNG Model \citep{weinberger2017,pillepich2018} using the \textsc{Arepo} code \citep{springel2010}. Similar to the The300 project, there was no iterative approach when selecting the initial volumes. Similarly to the aforementioned The300, a number of clusters in the TNG-Cluster project suffer from boundary particles within the virial radius \citep[see discussion in][]{nelson2024a}. The resolution, on the other hand, is improved compared to The300 clusters (see \cref{sim:props}). Science results include a possible connection of ICM velocities to halo and galactic properties \citep{ayromlou2024}, a prediction for X-ray velocity dispersion measurements from Perseus \citep{truong2024}, cool-core and non-cool core fractions of these clusters \citep{lehle2024} as well a study on the hot CGM of cluster satellites and how it can be retained by its host and observed in the X-ray band \citep{rohr2024}. 
 
The PICO cluster project \citep[Berlok et al. in prep.,][]{tevlin2025} is a recent zoom-in campaign for clusters using the \textsc{Arepo}/TNG Model. It is based on an increased parent box which has a size of $1h^{-1}$Gpc and the zoom-in regions are increased so that there are no boundary particles within the virial radius of the clusters at $z=0$. It contains 25 clusters performed with the \textsc{Arepo-2} code and the TNG galaxy formation model. Its focus lies on studies of plasma physical processes in the ICM, such as anisotropic viscosity and conduction and cosmic rays. The simulations have been used so far to demonstrate an accelerated magnetic field amplification via dynamo processes through including galaxy formation physics, especially at higher redshifts \citep{tevlin2025}.  

To allow a more compact overview of these different campaigns for cluster zoom-in simulations, we list some general properties of the aforementioned studies in table \ref{sim:props}. Among the properties we list the size of the parent box, the mass of the galaxy clusters and the typical size of the region free from boundary particles. In addition, figure \ref{sim:overview} shows a slice through the initial condition as well as the same slice in the $z=0$ cluster for the subset of these studies where we could access these data. All figures are normalized to the same scale to better visualize the differences in the high-resolution volume extraction methods and the degree and spatial distribution of contamination by low-resolution particles. We also include an example from the cluster ICs of this work for comparison. We will introduce our methods for creating them in the following section.

%--------------------------------------------------------------------

\section{Methods: SLOW Cluster Initial Conditions}
\label{sec:methods}
As mentioned, the studies introduced above typically select the high resolution region by finding particles within a certain distance or in the Friends-of-Friends (FOF) group of the central structure at $z=0$. Some are using iterative approaches to improve the size of the volume, which subsequently remains uncontaminated. However, they all commonly start from a sphere with a certain multiple of the virial radius (in most cases) around the cluster at $z=0$ to trace back the volume for the high resolution region within the ICs. In this work we propose an extension to this procedure in order to capture the dynamics of the region more accurately: We run an N-body only version of the parent simulation to the far future in order to reliably predict which particles interact with the central structures during the simulation time. This method allows us to create ICs with far more stable high-resolution regions than the traditional approaches for defining the high-resolution volumes. These initial high-resolution volumes however, are compatible with most modern IC codes like the aforementioned MUSIC \citep{hahn2011b} or \textsc{Ginnungagap} \citep{2025arXiv251110353P} codes since they can be passed as a mask to these codes.

In order to save computational cost, we choose a slightly different approach to creating particle ICs: Instead of re-sampling for each Lagrangian volume separately, we scale the entire simulation volume globally to the desired resolution levels for creating a layered high-resolution region (see \cref{methods:layering}) using \textsc{Ginnungagap}. For each zoom-in region, we then select the particles from the required resolution level based on the mask created in the previous step. With this method we run \textsc{Ginnungagap} only $N_\mathrm{Levels}$ times instead of $N_\mathrm{regions}$ times, which allows for a quicker extension of the cluster set in the future.

As a test set for this method we employ the highest resolution version of the SLOW (Simulating the Local Web) simulation, which is based on the realization number 8 of CLONES \citep{sorce2018b}. The SLOW simulation reproduces the local large scale density field with all its peculiarities \citep{dolag2023a} and has the additional advantage that the zoom-in regions obtained this way contain the simulated analogues of some of the most well studied clusters in the local Universe (see \Cref{Table:Clusters}) and their specific environments. For a discussion of general properties of these local clusters and supercluster regions see \citet{hernandez-martinez2024a,seidel2025b}.

\subsection{Clairvoyant - The parent simulation}
The parent simulation for the IC generation is a pure N-Body version of the SLOW simulation at various resolution levels, including versions with $768^3$, $1536^3$ and $3072^3$ resolution elements. For tracing the structures we used the initial conditions with $3072^3$ particles in a 500 $h^{-1}$ Mpc side length box, giving a mass resolution of $m_{dm}=3.1\times 10^8h^{-1}M_\odot$.
The method of obtaining the initial conditions for the parent box is described in detail in previous works \citep{sorce2018b,dolag2023a}, thus we limit the discussion here to a brief overview: From the Cosmicflows-2 catalogue of distances and peculiar velocities \citep{tully2013a} which suffers from shot noise and intrinsic uncertainty, the local velocity field is reconstructed using a Wiener Filter technique of variance minimization \citep{zaroubi1995,zaroubi1999}. There are a number of sophisticated bias correction procedures that are applied both to the raw data and the reconstructed field, to maximize the efficacy of the reconstruction, for details refer to: \citet{sorce2015,sorce2017,sorce2018a,sorce2018b}. Finally, the constraints are applied to the initial conditions using the Hoffman-Ribak algorithm \citep{hoffman1991}.

We performed a standard gravity-only run using these constrained ICs. Instead of using $z=0$ as the reference point for the zoom-in region extraction, as is done usually, we let the simulation propagate past this point to a scale factor of $a=1000$. 
The N-Body forward run is performed using  \textsc{OpenGadget3} \citep[OpenGadget3 collaboration in prep., refer to][and references therein]{groth2023a}, based on \textsc{Gadget2} \citep{springel2005h}. It incorporates several updates including an optimized tree-walk and an improved domain decomposition \citep{ragagnin2016}. We apply the \textsc{Subfind} algorithm \citep{springel2001a,dolag2009b} on the fly to all simulation outputs to obtain the Friends-of-Friends (FOF) particle groups.  
All simulations we ran adopt a background cosmology according to \citet{planckcollaboration2016a} with $H_0=67.77$\,km\,s$^{-1}$\,Mpc$^{-1}$, $\Omega_{\rm m}=0.307115$, $\Omega_{\Lambda}=0.692885$, and $\Omega_{\rm b}=0.0480217$.

\subsection{Cluster target selection}
 Using the constrained SLOW simulation, we can focus the generation of initial conditions on a number of identified analogues of local galaxy clusters \citep{hernandez-martinez2024a} as well as their large-scale arrangements, the local superclusters \citep{seidel2025b}. This provides a unique opportunity to study these well-observed regions in greater detail than previously possible. We therefore prioritized the selection to firstly cover the supercluster regions introduced by \citet{seidel2025b} as the first target volumes for the test set. This set contains the Coma, Virgo, Perseus-Pisces, Shapley and the Hercules superclusters. In addition, we added regions of high interest, including the A3667-A3651 region, where an elongated X-ray filament was discovered recently \citep{dietl2024}, as well as the A2199/97 region which is another well-known cluster pair \citep[e.g.][]{rines2001}. Additional clusters were added to the set based on the XRISM target selection \citep{xrismscienceteam2022} for PV, A01 and A02, with the long-term goal to provide high-resolution counterparts of these clusters for one-to-one ICM velocity structure comparisons \citep{groth2025}. A large number of these clusters also overlap with the ACCEPT database \citep{baldi2014} providing exquisite X-ray coverage and also have a large overlap with clusters where deep optical images allow investigations of the intra cluster light (ICL) properties \citep{kluge2025}. Therefore our first set consists of 20 regions containing 30 of the cross-matched galaxy clusters. The details of this region and the associated clusters are listed in table \Cref{Table:Clusters}. We have also identified 20 more regions for which we will create the zoom-in initial conditions in the future.
\begin{table}[h]
\small
\centering
\begin{tabular}{|l|l|l|l|l|l|l|}
\hline
& \multicolumn{2}{c|}{Reference} & \multicolumn{4}{c|}{Simulated Properties} \\ \cline{2-7}
Name & $N_\mathrm{constr.}$ & $M_{500c}^{Obs}$ & $M_{500c}^{Sim}$ & $X_{SG}$ & $Y_{SG}$ & $Z_{SG}$ \\ \hline
Virgo & 305 & 4.8 & 6.5 & -3.6 & 10.4 & -1.6 \\ \hline
Perseus & 447 & & & & & \\
A426 & -- & -- & 5.1 & 59.2 & 2.2 & -24.6 \\ 
AWM7 & -- & -- & 0.5 & 60.1 & -6.7 & -23.2\\ \hline
Bridge & 4 & & & & & \\
A3667 & -- & 5.8 & 4.4 & -156.8 & -111.3 & 66.7 \\ 
A3651 & -- & 3.2 & 2.1 & -145.6 &  -108.3 &  74.8 \\ \hline
A2029 & 4 & 6.8 & 6.0 & -92.6 & 142.4 & 141.4 \\ \hline
A1795 & 0 & 4.5 & 1.8 & 6.3 & 201.7 & 28.2 \\ \hline
Clstr pair & 43 & & & & & \\
A2199 & -- & 2.1 & 3.0 & -13.0 & 48.8 & 97.1\\ 
A2197 & -- &     &     &       &      &     \\ \hline
Hydra & 1859 & & & & &  \\ 
A1060 & -- & -- & 2.6 & -21.2 & 9.9 & -19.4 \\
A3526 & -- & 1.3 & 6.2 & -22.7 & 11.3 & -13.2 \\ \hline
A754 & 1 & 6.7 & 1.9 & -34.3 & 78.5 & -144.3 \\ \hline
A085 & 3 & 6.1 & 6.6 & 50.8 & -143.1 & 25.9 \\ \hline
A3395 & 49 & 3.0 & 7.0 & -53.3 & -43.2 & -101.7 \\ \hline
Shapley & 22 & & & & & \\ 
A3558 & -- & 4.4 & 9.3 & -141.0 & 66.1 & -37.2 \\
A3571 & -- & 4.7 & 4.0 & -142.4 & 54.1 & -33.9 \\
A3560 & -- & -- & 3.2 & -145.5 & 60.6 & -39.7 \\ 
A1736 & -- & -- & 3.3 & -137.5 & 65.6 & -22.6 \\ 
A3559 & -- & -- & 1.6 & -131.1 & 32.9 & -59.9 \\ \hline
A496 & 2 & 4.3 & 4.2 & -13.3 & -94.5 & -83.5 \\ \hline
Hercules & 30 & & & & & \\ 
A2147 & -- & 2.5 & 2.8 & -77.42 & 77.33 & 100.6 \\
A2151 & -- & 1.4 & 1.2 & -77.4 & 69.1 & 100.6 \\
A2152 & -- & 0.6 & 0.4 & -76.5 & 79.8 & 101.6 \\ \hline
Coma & 972 & & & & & \\
A1656 & -- & 5.3 & 9.5 & -2.9 & 82.3 & -9.1 \\
A1367 & -- & 1.8 & 1.1 & -2.9 & 71.4 & -18.7\\ \hline
Fornax & 32 & --- & 0.4 & 4.5 & -6.6 & -21.5 \\ \hline
A2256 & 3 & 6.2 & 1.7 & 115.9 & 96.3 & 79.4 \\ \hline
A119 & 191 & 3.4 & 5.8 & 73.6 & -117.3 & -24.9 \\ \hline
A644 & 2 & 5.1 & 5.0 & -26.5 & 70.6 & -204.2 \\ \hline
A1644 & 491 & 3.7 & 4.4 & -97.3 & 64.9 & -5.7 \\ \hline
A3266 & 9 & 6.6 & 8.6 & -60.0 & -93.2 & -128.1 \\ \hline
\end{tabular}
\caption{Overview of galaxy clusters in the IC set. $N_\mathrm{constr.}$ indicates the number of galaxy velocities from \citep{tully2013a} within the convex hull of the zoom-in region. This provides an estimate of how many tracers constrain the density field in the given zoom-in region. The observational mass estimates were obtained from the Planck cluster database PSZ1v2.1. We additionally list general simulated properties like masses and position taken from the parent box. Masses are given in units of $10^{14}M_\odot$, while positions are in cMpc/h.}
\label{Table:Clusters}
\end{table}
\subsection{Constructing the high-resolution Volume}
The common approach for defining a region of interest to re-simulate with higher resolution is to select the halo of interest at $z=0$ and define all particles in the simulation within a virial radius or multiples of the virial radius as the target Lagrangian region. These particles can then be traced back to the ICs of the box and there define the high-resolution volume, where subsequently the resolution is enhanced with one of the methods described below. The size one sets for the initial selection of the Lagrangian region aims to ensure that flows between the low-resolution volume and the high-resolution volume are minimized and at the same time keep the volume as small as feasible to reduce computational cost. To optimize this selection process, we propose an alternative approach for this pre-selection of the Lagrangian volume. The method is shape-free in the pre-selection process (as opposed to the spherical radii being used in previous zoom-in projects) and additionally incorporates dynamical information from each specific region. As a consequence the boundaries do not deform significantly, which ensures that boundary particles mix as little as possible with high-resolution particles. 

By letting the gravitational part of the parent SLOW simulation propagate into the far future ($a=1000$, as described in Paper I), we are able to extract a finally collapsed halo for a given structure by running the \textsc{Subfind} algorithm on the final snapshot output instead of the state of the simulation at $z=0$. Contained within this "final halo" are all dark matter particles that are gravitationally bound to the region (it was shown in Paper I, that the growth of massive haloes in fact stops before $a=1000$, we simply take the final snapshot of the forward simulation as the reference snapshot). To further ensure boundary stability we incorporate the local residual particle velocities to extend this Lagrangian volume adaptively. The bounding surfaces are computed using a HealPIX tesselation \citep{gorski2005} of a sphere centered on the target halo. Since nearby, but non-merging massive structures can impact the stability of such a boundary significantly through tidal interactions, we hierarchically merge adjacent Lagrangian Volumes of galaxy clusters with the volume of interest.

\begin{figure}[ht]
   \centering
   \includegraphics[width=0.49\textwidth]{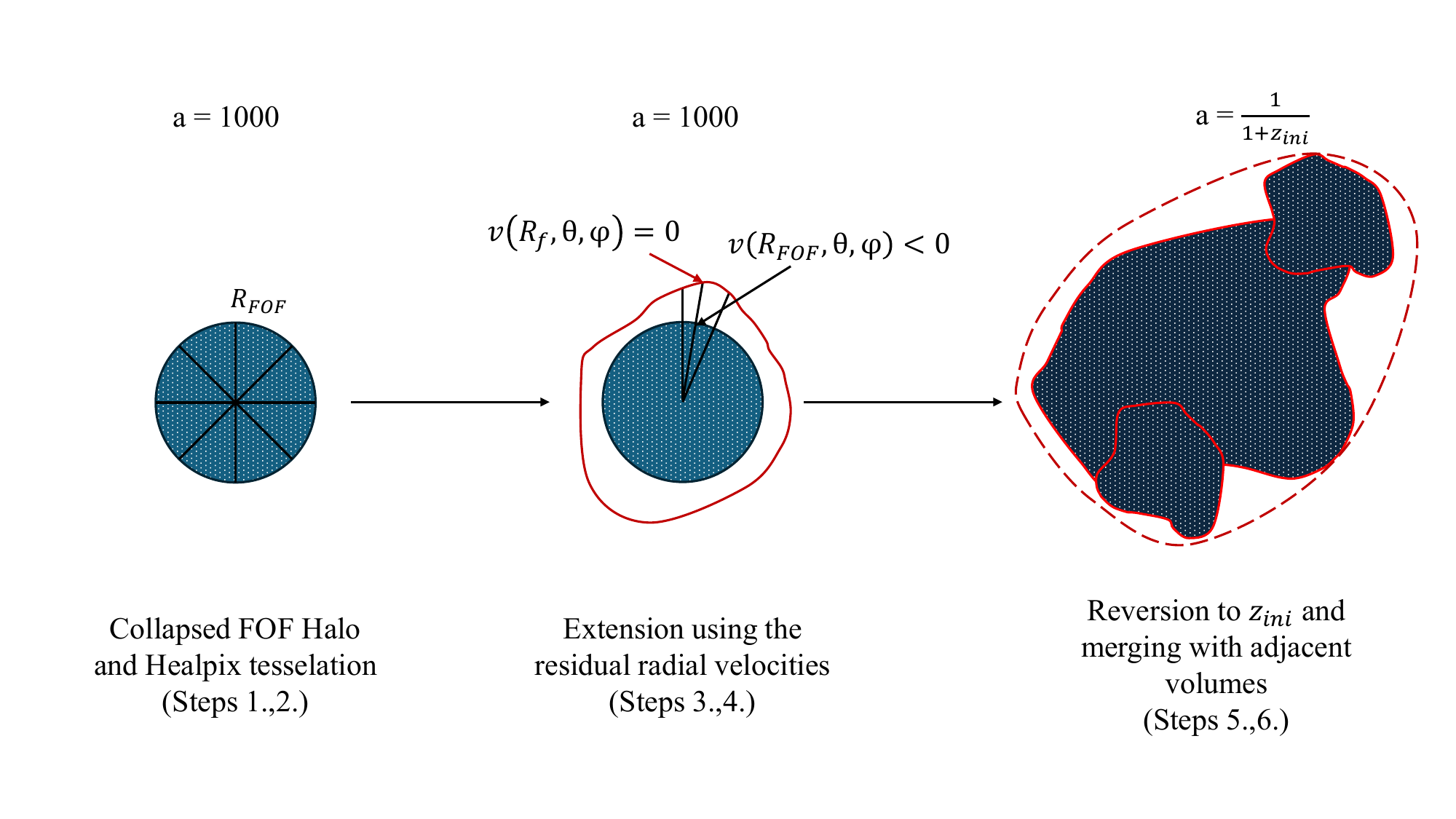}
   \caption{Schematic representation of the boundary creation process. Note that the collapsed FOF groups are not necessarily spherical, but are drawn as such for simplicity.}
              \label{pic:schematic}
\end{figure}
The complete process to compute the Lagrangian volume is then as follows:
\begin{enumerate}
\item{Extract FOF group from the snapshot at $a=1000$}
\item{Create a Healpix tessellated distance map of these particles, where each angular pixel is mapped to the largest FOF particle distance within the angular bin}
\item{Starting at this distance for each pixel, compute the radial velocity}
\item{For each pixel with a negative (inward) radial velocity, compute the radial velocity profile and set as the new pixel value the distance at which the radial velocity crosses zero.}
\item{Repeat this process for close by massive structures\footnote{This step depends on the volume of interest and how large one wants to allow the final region to be. For studies focusing only on the central cluster of a given region, computational trade-offs can be made by not considering adjacent structures.}}
\item{Find all particles that lie within any of the resulting bounding surfaces at the starting redshift $z_\mathrm{ini}=120$}
\end{enumerate}
 \cref{pic:schematic} visualizes this process as a schematic drawing.

Tracing the set of particles obtained with this method back to the ICs of the forward run, we obtain the final selection volume by once again putting the particles on a HealPIX distance map and setting the maximal radial distance within each pixel as the boundary radius in the corresponding angular bin. To obtain a smoother boundary we additionally upscale this map to a higher resolution and smooth over it with an angular gaussian kernel.

\subsection{Multimass layering}
\label{methods:layering}
\begin{figure}[ht]
   \centering
   \includegraphics[width=\columnwidth]{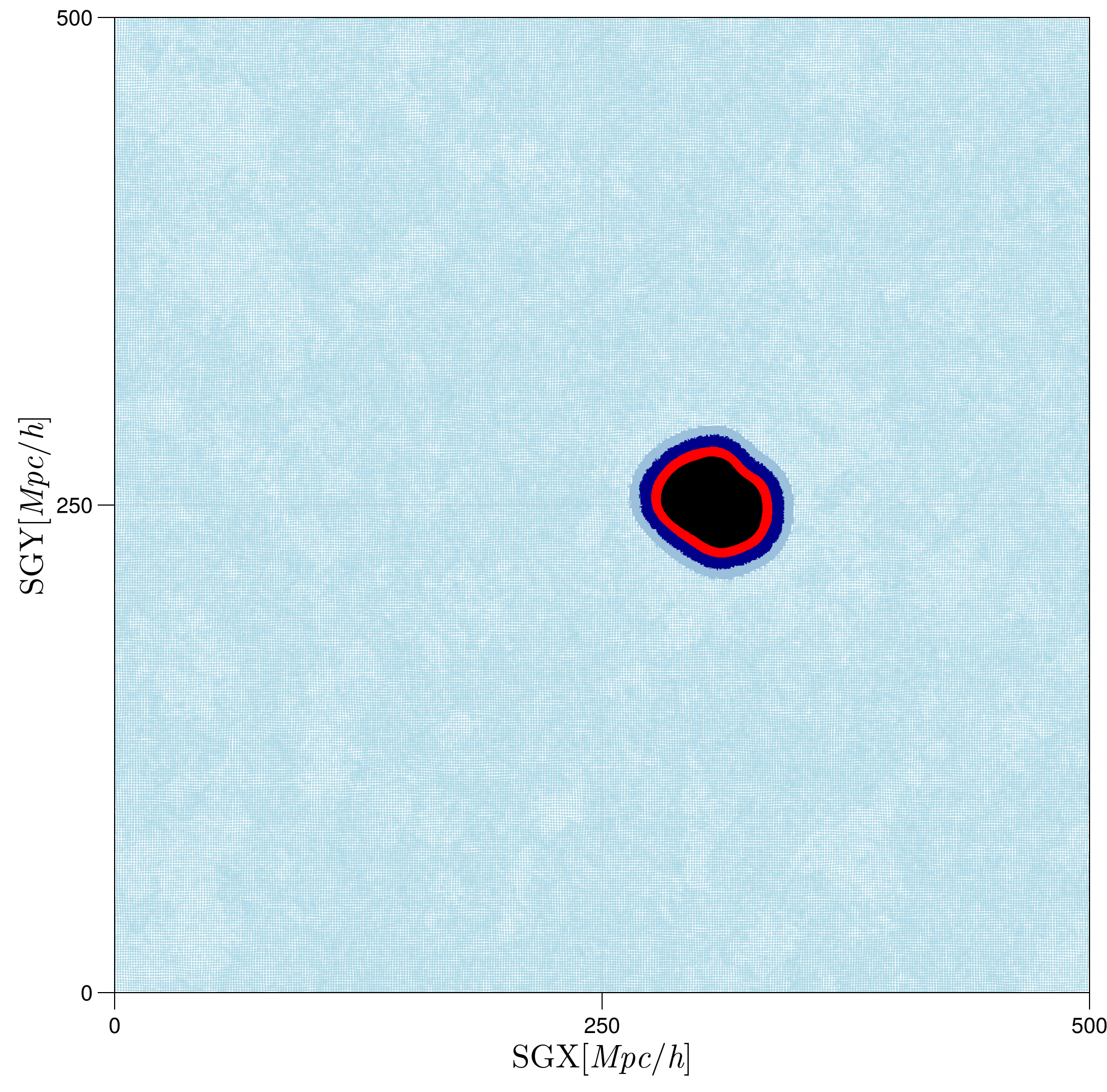}
   \caption{Layered ICs from the SLOW Perseus counterpart with five resolution levels. Each color represents a different mass resolution, resolution decreases with increasing radius.}
              \label{PIC}%
    \end{figure}

There are several methods to combine volumes of different resolution into one simulation. The approach we use in this work is comparable to the multimass approach employed by the ZIC code \citep{tormen1997}: We start by creating several instances of the parent box in different resolution levels using the GINNUNGAGAP\footnote{https://github.com/ginnungagapgroup/ginnungagap} code \citep{2025arXiv251110353P}. Particles within the high-resolution selection volume are extracted from the highest resolution box. We then add several boundary layers (depending on the target resolution) by upscaling the selection volume self-similarly several times and selecting the particles between the boundary surfaces from ICs of decreasing resolution. An example of such a region and its enclosing can be seen in figure \ref{PIC}.  The current highest resolution grid we are using is the $6144^3$ realization of the box. Additionally we keep the $3072^3$, $1536^3$, $768^3$ and $384^3$ realizations as low resolution version of the full box for boundary selection and lower resolution test zoom-ins. This allows for a combination of five layers of decreasing resolution ($6144^3-384^3$) for the highest resolution zoom-ins  and, correspondingly, four layers for the second highest ($3072^2-384^3$). We thus call these the L5 and L4 sets respectively with the L5 being the "gold standard" set. 

This layering dampens possible spurious effects from very high-mass and low-mass particles interacting. We ensure mass conservation by checking first if a given particle of the low resolution grid lies within the applied boundary and obtaining its children from the high resolution grid if this is the case. The fact that particles can be obtained self-consistently from the different full-box resolution grids is a major advantage of pre-computing the box in all applied resolutions beforehand instead of increasing the resolution exclusively in the designated volume. In the future, we plan to extend the current range to even higher resolutions for the zoom-in regions as well as lower resolution versions based on the $3072^3$ resolution for testing. Both operations are feasible within the GINNUNGAGAP framework. A drawback of this approach is that it is relatively expensive in terms of memory, to create the high-resolution grid, therefore it can be considered more as a lateral approach to quickly add many regions to the sample from a given high-resolution volume.

\begin{figure}[ht]
   \centering
   \includegraphics[width=0.49\textwidth]{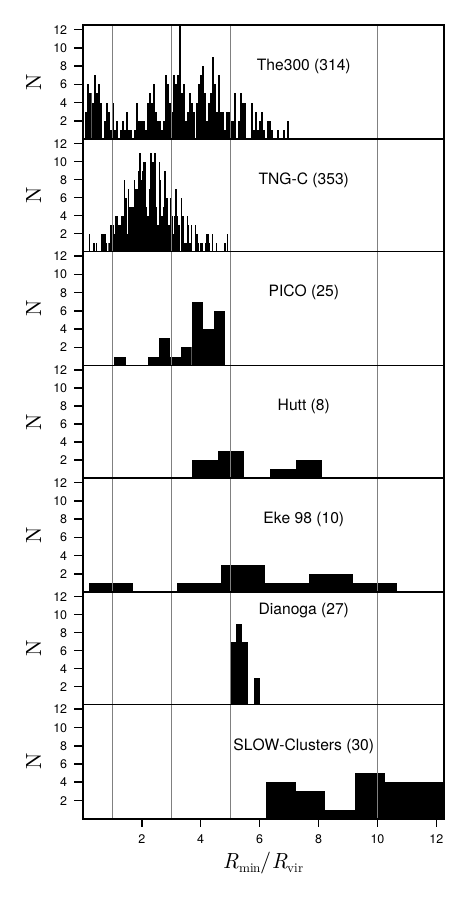}
   \caption{Distribution of the minimum encounter distance relative to the virial radius for the different IC sets. Gray lines mark 1,3,5 and 10 $R_{\mathrm vir}$ as also indicated as circles in figure \ref{sim:overview}.}
              \label{pic:encounter}
\end{figure}

\section{Results}
We ran the dark-matter-only version of the ICs generated with the method introduced above. The resulting structures at $z=0$ can be seen in \Cref{pic:lower_decks_full} for 30 local galaxy clusters found within the Lagrangian region set. As can be seen from the figure some of these regions contain multiple local cluster analogues because they are actively merging supercluster regions at $z=0$ (for further details see \citet{seidel2025b}). As can already be gathered from the gray circles in \Cref{pic:lower_decks_full} as well as the boundary distances shown in the lower left panels, the ICs provide an unprecedented level of cleanliness, with boundary particles at least $\sim 6r_\mathrm{vir}$, in some cases even far beyond $\sim 10\,r_\mathrm{vir}$, at a large distance from any of the clusters of interest. 
%We will now analyze this in more detail.

\subsection{Cleanliness}
\Cref{sim:overview} visualizes the evolution of zoom-in ICs including the SLOW-C set created with the aforementioned methods. As can be seen from the top row of the figure, the high resolution and boundary regions of the SLOW cluster ICs are comparable in size to their historical predecessors.
In the bottom row, showing the final state at $z=0$, however it becomes apparent that the adaptive nature of the SLOW cluster set, taking into account the whole collapsing region around the central structure, helps the boundary to be much more stable across time. In the final snapshot there is no boundary particle (shown in red) within 10 virial radii of the cluster of interest, a clean region extent that none of the comparison ICs can reproduce. For a more quantitative comparison between previous simulations and our ICs, we show the minimum encounter distance distribution for various comparison sets and all the 30 SLOW cluster from our 20 selected volumes in \Cref{pic:encounter}. The lowest SLOW encounter distances are comparable to or exceed most of the other simulation suites' highest encounter distances. This shows that the SLOW cluster IC achieve unprecedented cleanliness, which will allow a much larger region to be studied from a hydrodynamical perspective and ensures that the central cluster or clusters are absolutely free of spurious effects caused by low resolution dark matter particles. This additionally allows to properly study the immediate environment of the cluster, including the filaments which are connecting the cluster to the cosmic web.

\subsection{Mixing layer}
\begin{figure}[ht]
   \centering
   \includegraphics[width=0.49\textwidth]{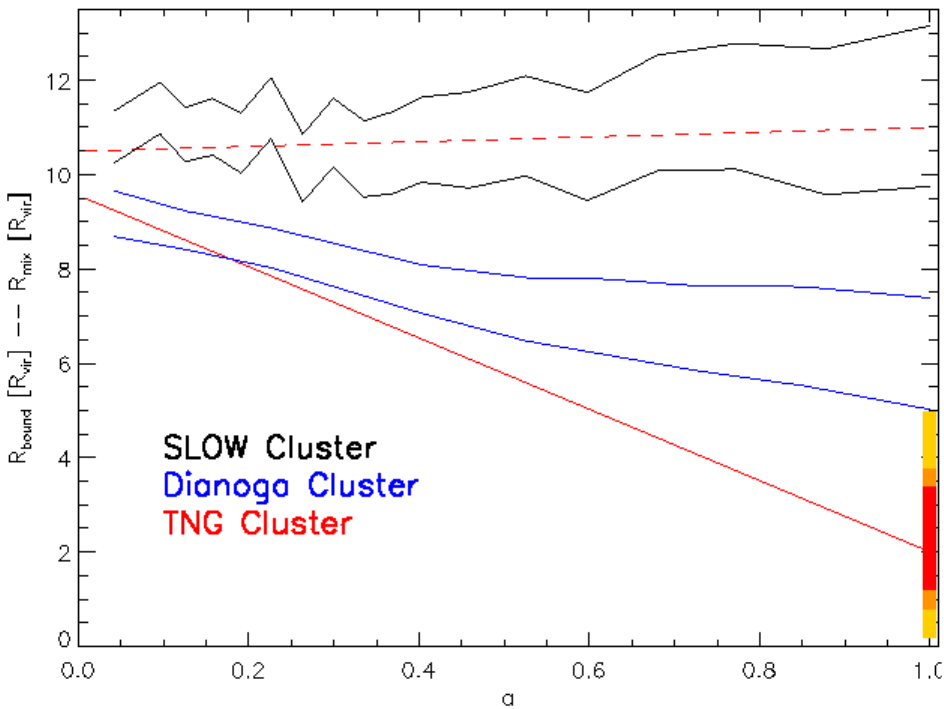}
   \caption{Evolution of the boundary layer for the SLOW cluster (black), Dianoga cluster (blue) and TNG Cluster (red), always expressed in units of the virial radius of the cluster. The lower line indicates the minimum distance of a boundary particle to the cluster center. For the SLOW/Dianoga cluster a typical cluster is shown to highlight the individual evolution of such a system. The full distribution of SLOW-C and Dianoga at $z=0$ can be read off from \cref{pic:encounter}. For the TNG clusters the median value is shown and the histogram at $a=1$ (i.e. $z=0$) shows the distribution, where the yellow region shows 10\% of the cluster population, the orange 20\% and the red the remaining 70\%. The upper lines show the maximum distance of high resolution particles, so together with the lower line they mark the mixing layer between boundary and high resolution particles, where physical processes can be significantly influenced by this mixing.}
              \label{pic:boundaries}
\end{figure}

\Cref{pic:boundaries} shows the co-evolution of the minimal encounter distance for low-resolution particles and the maximal encounter distance for high-resolution particles for one of the SLOW clusters and two clusters from other IC sets for comparison. Effectively these two curves give the evolution of the mixing layer, where high-resolution particles and low-resolution particles co-exist. The evolution of the SLOW cluster shows the mixing layer to be relatively stable for this set: Both the minimal boundary distance and the mixing layer stay almost constant from $a=0$ to $a=1$ for the SLOW cluster. For the Dianoga set, where the target size of the clean regions was set to $\approx 5r_{\mathrm{vir}}$, the depth of the mixing layer is only slightly increasing. Nevertheless the clean region is significantly shrinking as the initial cutout is still within the in-fall region of the final structure. In addition, we added the reported behavior of the initial and final size of the sphere of boundary particles \citep{nelson2024a} as well as its distribution for all TNG-Clusters as colored histogram at $z=0$. The reason why the contamination in the TNG-Clusters is significantly higher than in the Dianoga clusters, despite starting from similarly sized Lagrangian Volumes is the fact that the creation of the Dianoga cluster used several iterations to obtain a optimized description of the shape of the region in the ICs. 

\begin{figure}[ht]
   \centering
   \includegraphics[width=0.49\textwidth]{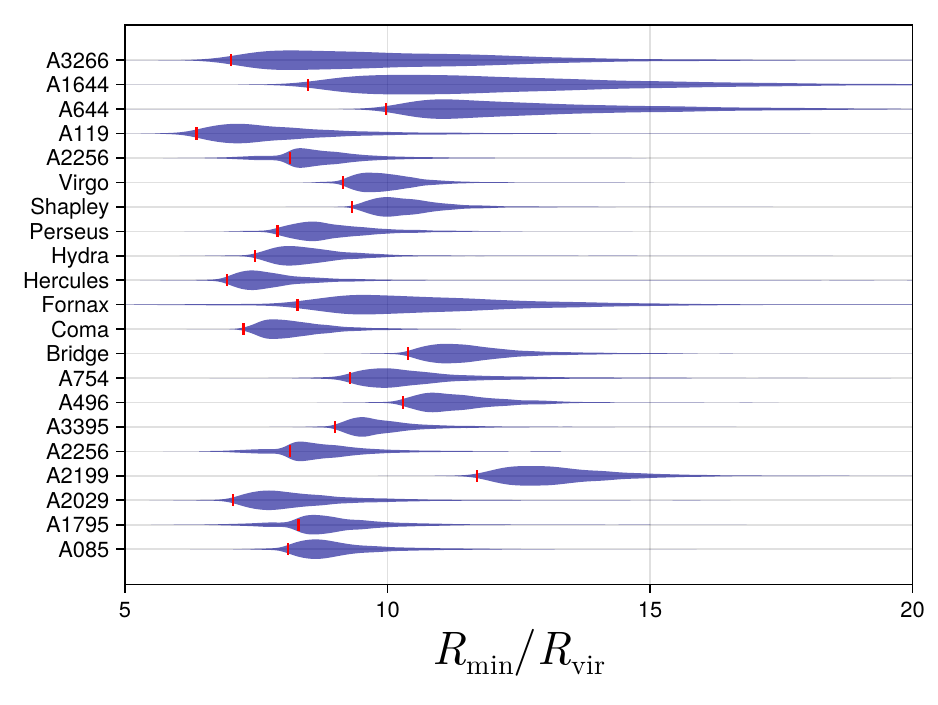}
   \caption{Minimal boundary distances for the individual regions from the SLOW cluster set (red vertical lines) at $z=0$. The dark blue violins additionally show the histogram of mixing layer thickness $D+r_\mathrm{min}$ across an NSIDE=16 healpix map centered on the main cluster. Values $<r_\mathrm{min}$ therefore indicate a negative thickness, i.e. the high resolution particles and the low resolution particles are separated at $z=0$.}
              \label{pic:pixhist}
\end{figure}

Because the regions from the SLOW IC set can be highly aspherical we additionally show the angular distribution of mixing layer thickness for the 20 simulation volumes, centering on the most massive cluster in \Cref{pic:pixhist}. This statistic is computed by putting the high-resolution particles and the boundary particles on a healpix map (NSIDE=16) and then subtracting the minimal boundary particle distance from the center from the maximal high-resolution particle distance. For comparability we added the distances $D$ to the global minimal boundary particle distance $r_\mathrm{min}$ for each cluster, so the actual maximal high-resolution particle distances are not reflected by this statistic. From the figure we see that for all of the clusters, the mixing layer is small compared with the overall size of the clean region. Some of the clusters additionally exhibit a separation region, where there are no high resolution particles overlapping the boundaries (shown by the tails of the histograms that lie to the left of the minimal boundary distance). 


\section{Conclusions}
We introduced a novel method for constructing zoom-in ICs achieving high-resolution regions that are free of contamination from low-resolution particles to an unprecedented degree.
To evaluate this method, we constructed a suite of regions encompassing prominent galaxy clusters from the constrained SLOW simulation. In total we evolved 20 such regions, containing 30 cross-identified clusters, utilizing two resolution levels until $z=0$  to evaluate for their cleanliness. We find the following:
   \begin{enumerate}
      \item The boundaries, which are composed of layers of particles of decreasing resolution (and conversely increasing mass) are very stable across time, showing little deformation from the initial shape at any given redshift.
      \item All 30 clusters are free of any boundary particles to within at least 6$r_\mathrm{vir}$, half of them have clean volumes even beyond $\approx 9r_\mathrm{vir}$. This significantly improves on previous zoom-in initial condition suites.
      \item There is a certain degree of high-resolution particles diffusing outwards, however these so-called mixing layers are small compared to the size of the pristine volume ($\approx 10 \%$)
   \end{enumerate}
Due to these properties, our method is ideally suited for carrying out high-fidelity hydrodynamical simulations including the large scale cluster environment and the filaments which connect the clusters to the cosmic web. In future work we will introduce a set of fully hydrodynamical simulations of the set presented in this paper. This will be the first large set of such re-simulations built on constrained counterparts of real local Universe clusters. With the mass resolution achieved, we will also be able to study galaxy formation in these realistic large scale environments, expanding on previous work \citep[e.g.]{sorce2021,libeskind2020,gallagher2026}.  Due to the large uncontaminated volume encompassing these clusters, this additionally makes it possible to study the large scale environmental effects of galaxy clusters and their surrounding filaments on galaxy formation in great detail. 
These advantages make these simulations invaluable for interpreting the results from ongoing observational campaigns. We plan to generate detailed X-Ray mock observations and spectra in order to provide theoretical predictions and comparisons for the ICM dynamics and morphology as observed by Chandra, XMM and XRISM, as well as through the eROSITA all-sky survey for these local clusters. As the large-scale environment of our clusters is well-constrained, we will also be able to carry out novel galaxy population studies in the future, expanding on the work which has been already done on the Coma \citep{malavasi2023,zoller2025a} and Virgo \citep{sorce2021,lebeau2024,sorce2026} cluster twins.  
\begin{acknowledgements}
This work was supported by the grant agreements ANR-21-CE31-0019 / 490702358 from the French Agence Nationale de la Recherche / DFG for the LOCALIZATION project. KD acknowledges support by the COMPLEX project from the European Research Council (ERC) under the European Union’s Horizon 2020 research and innovation program grant agreement ERC-2019-AdG 882679. The calculations for the simulations were carried out at the Leibniz Supercomputer Center (LRZ) under the project pn68na. JS is supported by the University of Lille via the Welcoming Internationals to Lille initiative for the UNIVERSITWINS project.
\end{acknowledgements}

\bibliographystyle{aa} % style aa.bst
\bibliography{ICPaper.bib}
\begin{appendix}
\section{The SLOW cluster zoom-IC suite}
\begin{figure*}[ht!]
   \centering
   \includegraphics[width=0.99\textwidth]{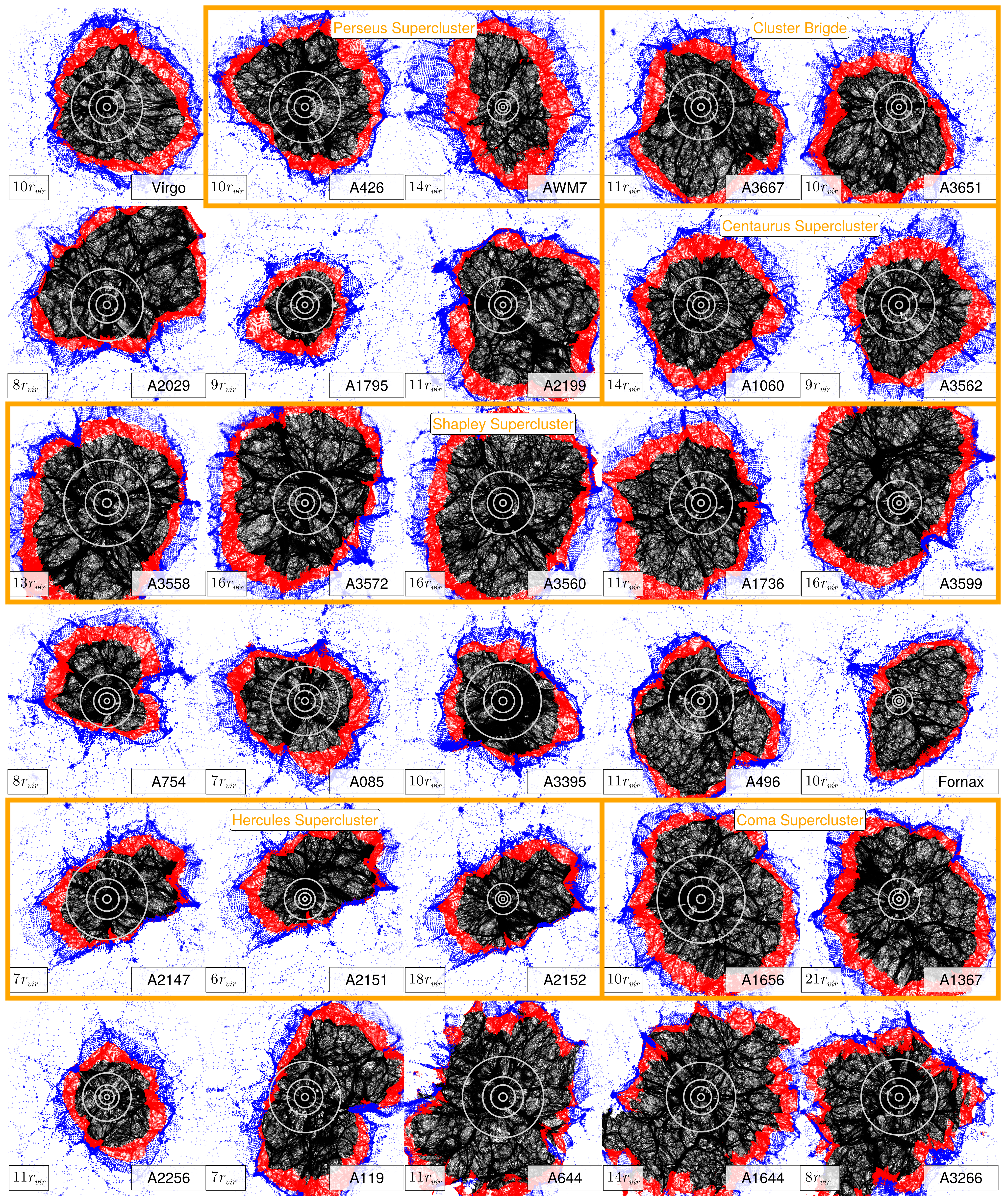}
   \caption{Thin slice of one cMpc/h through the different galaxy clusters belonging to the first set of the SLOW Cluster selection. Shown is a 100 cMpc/h region with the high-resolution particles scattered in black, the first boundary in red and subsequent outer boundaries in blue with opacity decreasing with increasing distance from the central cluster. The supercluster regions containing multiple clusters are indicated with the orange frames. The gray circles indicate 1,3,5 and 10 times the virial radius of the respective cluster. }
              \label{pic:lower_decks_full}
\end{figure*}

As mentioned in previous sections the full SLOW cluster set contains 30 local galaxy clusters in 20 regions. We show the full sample at the final snapshot in \cref{pic:lower_decks_full}. Consistent with previous figures, the regions can be highly aspherical, nevertheless they remain free of boundary particles out to large distances from the central structure. For the regions containing multiple clusters (e.g. the Shapley supercluster region) it is apparent that, while the overall region is centered on the dominant structure, the secondary clusters stay similarly uncontaminated as they belong to the same basin of influence and can thus be cleanly isolated from the surrounding box. 
\end{appendix}
\end{document}